\documentclass[12pt,preprint]{aastex}
\usepackage{epsf}
\usepackage{color}

\newcommand{\eqn}[1]{Eq.\,(\ref{#1})}

\newcommand{\be}{\begin{displaymath}}
\newcommand{\ee}{\end{displaymath}}
\newcommand{\bea}{\begin{eqnarray}}
\newcommand{\eea}{\end{eqnarray}}

\newcommand{\kap}[1]{Section \ref{#1}}
\newcommand{\mem}[1]{\ensuremath{\mathrm{ #1}}}
\newcommand{\msun}{\ensuremath{M_\odot}}

\newcommand{\abb}[1]{Fig.\,\ref{#1}}
\newcommand{\czw}{\ensuremath{^{12}\mem{C}}}
\newcommand{\ose}{\ensuremath{^{16}\mem{O}}}
\newcommand{\nadr}{\ensuremath{^{23}\mem{Na}}}
\newcommand{\mgvi}{\ensuremath{^{24}\mem{Mg}}}
\newcommand{\nezwa}{\ensuremath{^{20}\mem{Ne}}}

\usepackage{natbib}

\shortauthors{Denissenkov et al.}
\shorttitle{CBM and Hybrid C/O/Ne White Dwarfs}

\begin{document}
\title{The C-flame Quenching by Convective Boundary Mixing in Super-AGB Stars and
the Formation of Hybrid C/O/Ne White Dwarfs and SN Progenitors}

\author{P. A. Denissenkov\altaffilmark{1,2}, F. Herwig\altaffilmark{1,2}, 
        J. W. Truran\altaffilmark{2,3}, and B. Paxton\altaffilmark{4}}
\altaffiltext{1}{Department of Physics \& Astronomy, University of Victoria,
       P.O.~Box 3055, Victoria, B.C., V8W~3P6, Canada,
       pavelden@uvic.ca, fherwig@uvic.ca}
\altaffiltext{2}{The Joint Institute for Nuclear Astrophysics, Notre Dame, IN 46556, USA}
\altaffiltext{3}{Department of Astronomy and Astrophysics, and Enrico Fermi Institute, University of Chicago, Chicago, IL 60637 USA}
\altaffiltext{4}{Kavli Institute for Theoretical Physics and Department of Physics, Kohn Hall, University of California, Santa Barbara, CA 93106, USA}

\begin{abstract}
After off-center C ignition in the cores of super-AGB stars the C flame
propagates all the way down to the center, trailing behind it the
C-shell convective zone, and thus building a degenerate ONe core. This
standard picture is obtained in stellar evolution simulations if the
bottom C-shell convection boundary is assumed to be a discontinuity
associated with a strict interpretation of the Schwarzschild condition
for convective instability. However, this boundary is prone to
additional mixing processes, such as thermohaline convection and
convective boundary mixing. Using hydrodynamic simulations we show that,
contrary to previous results, thermohaline mixing is too inefficient to
interfere with the C-flame propagation. However, even a small amount of
convective boundary mixing removes the physical conditions required for
the C-flame propagation all the way to the center. This result holds
even if we allow for some turbulent heat transport in the CBM region. As
a result, super AGB stars build in their interiors hybrid C-O-Ne
degenerate cores composed of a relatively large CO core
($M_\mathrm{CO}\approx 0.2 \msun$) surrounded by a thick ONe zone
($\Delta M_\mathrm{ONe}\ga 0.85 \msun$) with another thin CO layer
above. If exposed by mass loss, these cores will become hybrid C-O-Ne
white dwarfs. Otherwise, the ignition of C-rich material in the central
core, surrounded by the thick ONe zone, may trigger a thermonuclear
supernova explosion. The quenching of the C-flame may have implications
for the ignition mechanism of SN Ia in the double-degenerate merger
scenario. 
\end{abstract}


\section{Introduction}

The evolution of a single star is determined by its initial chemical
composition and mass. In this paper, we consider models of super
asymptotic giant branch (SAGB) stars of compositions close to solar. By
definition, the initial masses of SAGB stars have to be sufficiently
high ($M > M_\mathrm{up}$) to burn C in their hydrogen and helium
exhausted cores, while being too low ($M < M_\mathrm{mas}$) to ignite
neon afterwards \citep{garcia-berro:94}. The main products of C burning
are oxygen and neon, therefore the SAGB stars build ONe degenerate cores
in their interiors. What will eventually happen with the ONe core
depends on the ratio of its growth rate and the stellar mass-loss rate 
\citep{poelarends:08}. Like in their less massive AGB counterparts, the
cores of SAGB stars can increase in mass because they are surrounded by
H and He burning shells. The latter intermittently experiences thermal
pulses, possibly followed by dredge-up,  after having accumulated a
certain amount of fresh He from the overlying H shell. One of the major
uncertainties in the core growth rate comes from the poorly constrained
depth of the He shell that is reached by the base of the surface
convection zone immediately after a He thermal pulse, during the third
dredge-up. The details of the cumulative core-growth depend on the
convective boundary mixing assumptions for the bottom boundary of the
convective envelope that may or may not trigger the third-dredge up. If
the core mass reaches the Chandrasekhar limit $M_\mathrm{Ch}\approx 1.37
M_\odot$ before the star has lost its H-rich envelope then
electron-capture reactions, starting on Mg and Ne, will trigger its
collapse leading to an electron-capture supernova \citep{miyaji:80}.
Otherwise, an ONe white dwarf will be born \citep{pumo:09}.

The neutrino energy losses in high-density CO degenerate cores,
predominantly via the photo and plasma neutrinos, cause a temperature
inversion with the maximum located some distance from the center.
Therefore, just as the He-core flash, C burning in SAGB stars starts
off-center if the initial mass is not too close to $M_\mathrm{mas}$.
Because of the degenerate conditions, the first episode of C ignition
takes the form of a thermal flash that slightly reduces C abundance in
the C-shell convective zone and eventually dies out. The following
phase of C burning occurs near the bottom of the extinct convective zone
under less degenerate conditions. This time, the C flame propagates all
the way toward the center in the form of deflagration, pulling along the
convective zone (\abb{fig:f1}). However, this is true only for
the case when all other possible mixing processes, except convection
operating within Schwarzschild boundaries, are neglected
\citep{siess:06}. The situation radically changes when thermohaline
mixing is added below the bottom of the C-shell convective zone,
where the mean molecular weight increases with the radius as a result of
the off-center C burning. \cite{siess:09} has shown that the C flame
stops propagating toward the center in this case because thermohaline
mixing deprives the C-flame precursor of fuel. We can reproduce this
behaviour (\abb{fig:f2}) when making the same assumption on the
efficiency of thermohaline mixing (see below). In that case, by the end
of C burning as much as 2\,--\,5\% of unburnt C (in mass fraction)
remains inside the ONe core which, according to \cite{siess:09}, may
modify its electron-capture induced collapse later on.

However, the efficiency of thermohaline mixing has been since
investigated in more detail. When it starts, thermohaline mixing has a
pattern of rising and sinking fluid parcels that resemble fingers. The
latter are often called ``salt fingers'' because it is an unstable
salinity distribution that causes their growth in the ocean. In
one-dimensional computations, the efficiency of thermohaline mixing can
be parameterized using the salt-finger aspect ratio $a = l/d$, where $l$
and $d$ are finger's length and diameter \citep{denissenkov:10}. In his
simulations, \cite{siess:09} has actually used a value of $a\approx 7$
needed to explain the observed evolutionary declines of the surface C
abundance and $^{12}$C/$^{13}$C ratio in low-mass stars on the red giant
branch (RGB) above the bump luminosity, as proposed by
\cite{charbonnel:07}. However, the recent two- and three-dimensional
numerical simulations of thermohaline convection in a low-mass RGB star
have independently produced an estimate of $a < 1$
\citep{denissenkov:10,traxler:11,denissenkov:11}. We will show that this
estimate is still valid for the CO cores of SAGB stars and that it
renders thermohaline mixing too inefficient to prevent the C flame from
reaching the center.

Does this leave us where we were before the work of \cite{siess:09}? Our
answer is no, because another mixing process relevant to the convective
C-shell burning in the cores of SAGB stars has yet to be accounted for.
Indeed, it is known that at the boundaries of shell-flash convection,
such as those in classical novae or in AGB stars, shear motion induced
by convective flows and internal gravity waves lead to mixing beyond the
Schwarzschild convective boundaries, in which the Kelvin-Helmholtz
instability plays an important role
\citep{glasner:97,herwig:06,casanova:11}. The most relevant, because
similar, case is such convective boundary mixing (hereafter, CBM) at the
bottom of the He-shell flash (or pulse-driven) convection zone in AGB
stars \citep[e.g.][]{herwig:99,miller-bertolami:06,weiss:09}.  The
consequences include larger \czw\ and \ose\ abundances in the
intershell, in agreement with observations of H-deficient post-AGB stars
\citep{werner:06}. We have included CBM in our simulations of C burning
in the cores of SAGB stars using a prescription that is supported by the
recent multi-dimensional hydrodynamic simulations of He-shell flash
convection \citep{herwig:06,herwig:07}. We have found that CBM, like
thermohaline convection with $a=7$ and $a=10$, forces the C flame to
stop propagating toward the center. The main difference between the two
mixing processes is that CBM is present only in the vicinity of
convective boundaries, therefore C is left unburnt in the entire core
below a narrow region adjacent to the bottom of the convective zone.
This leads to a new evolutionary path in which the SAGB stars, at least
those with the initial masses not too close to $M_\mathrm{mas}$, build
in their interiors degenerate cores composed of a relatively large CO
core surrounded by a thick ONe zone with a thin CO layer on the top.
Such hybrid C-O-Ne degenerate cores will behave differently when
the C-rich material in their central parts, surrounded by thick ONe
zones, eventually ignites, or when they become white dwarfs and begin
cooling down.

In \kap{sec:no_mix}, we analyze the physics of C-flame propagation
toward the center in the absence of extra mixing processes and identify
its most important driving mechanism. In \kap{sec:ovsh}, we explain why
this mechanism does not work when CBM is taken into account. In
\kap{sec:th}, we present the results of our hydrodynamic simulations
of thermohaline convection in a region below the C-flame convection zone
and apply them  to simulate the C-flame propagation in an SAGB star. The
effect of heat transport in the CBM algorithm is investigated in
\kap{sec:CBM_heat}, while \kap{sec:concl} contains discussion and
conclusions.

\section{The C-flame propagation in the absence of extra mixing}
\label{sec:no_mix}

We have computed the evolution of a star with the initial mass
$9.5\msun$  and metallicity  $Z=0.02$ from the pre-main sequence to C ignition in its
degenerate CO core using the state-of-the-art stellar evolution code of
MESA revision 4631 \citep{paxton:11,paxton:13}. Thus, our first SAGB
model has the same initial parameters as the corresponding model of
\cite{siess:06} with which we compare our results. We have used the MESA
equation of state (EOS). It adopts the 2005 update of the OPAL EOS
tables \citep{rogers:02} supplemented for lower temperatures and
densities by the SCVH EOS that accounts for partial dissociation and
ionization caused by pressure and temperature \citep{saumon:95}. 
Additionally, the HELM (\citealt{timmes:00}) and PC \citep{potekhin:10}
EOSs cover the regions where the first two EOSs are not applicable. 
There are smooth transitions between the four EOS tables. We have used
the OPAL opacities \citep{iglesias:93,iglesias:96} supplemented by the
low temperature opacities of \cite{ferguson:05}, and by the electron
conduction opacities of \cite{cassisi:07}. The nuclear network consists
of 31 isotopes from H to $^{28}$Si coupled by 60 reactions that account
for H burning in pp chains, CNO-, NeNa-, and MgAl-cycles, as well as He,
C, and Ne burning. By default, MESA uses reaction rates from
\cite{caughlan:88} and \cite{angulo:99}, with preference given to the
second source (NACRE). It includes updates to the NACRE rates for
$^{14}$N(p,$\gamma)^{15}$O (\citealt{imbriani:05}), the triple-$\alpha$
reaction \citep{fynbo:05}, $^{14}$N$(\alpha,\gamma)^{18}$F
\citep{gorres:00}, and $^{12}$C$(\alpha,\gamma)^{16}$O \citep{kunz:02}.

\abb{fig:f3} shows profiles of various stellar structure
variables at and near the bottom of the C-flame convection zone in the
Schwarzschild-only convective boundary model 4900 from \abb{fig:f1}. 
\cite{siess:06} has demonstrated
that, for the C flame to continue propagating all the way toward the
center, the maximum of the nuclear energy generation rate
$\varepsilon_\mathrm{nuc}$ should precede the maximum on the $\log T$
curve. The relative positions of the vertical solid and dashed lines in
the upper- and middle-left panels in \abb{fig:f3} confirm that
this is true for our simulation. In this case, the energy of C
burning released at the maximum of $\varepsilon_\mathrm{nuc}$ pre-heats
plasma in front of the maximum of $\log T$, thus facilitating its
advancement. \cite{siess:06} has also noted that the location of the
bottom of the C-flame convection zone is tightly bound with the location
of the maximum of $\log T$. Indeed, the former is determined by the
Schwarzschild criterion, $\nabla_\mathrm{rad} = \nabla_\mathrm{ad}$,
where
\bea
\nabla_\mathrm{rad} = \frac{3\kappa}{16\pi G ac}\frac{P}{T^4}\frac{L_r}{M_r},
\eea
and $\nabla_\mathrm{ad} = (\partial\ln T/\partial\ln P)_S$ are the radiative and adiabatic temperature gradients (logarithmic
and with respect to pressure), $\kappa$ is the opacity, other quantities having their usual meanings. 
Given that $\nabla_\mathrm{rad} = 0$
at the location of the maximum of $\log T$ because $L_r\propto dT/dr =0$ there, 
the Schwarzschild criterion is satisfied at some small distance above it, where
\bea
L_r = L_\mathrm{Sch} = \frac{16\pi G ac}{3\kappa}\frac{T^4}{P}M_r\nabla_\mathrm{ad}.
\label{eq:LSch}
\eea
The increase of the luminosity from its zero value at the point $M_r(T_\mathrm{max})$, where $\nabla_\mathrm{rad} = 0$, 
to its critical value $L_\mathrm{Sch}$
 at the convective boundary at $M_\mathrm{Sch}$, where $\nabla_\mathrm{rad} = \nabla_\mathrm{ad}$, 
is provided by the generation of C-burning nuclear energy in this mass interval,
\bea
L_\mathrm{Sch} = \int_{M_r(T_\mathrm{max})}^{M_\mathrm{Sch}}\varepsilon_\mathrm{nuc}dM_r.
\eea

For the maximum of $\varepsilon_\mathrm{nuc}$ to precede that of $\log T$, the derivative 
$(d\log\varepsilon_\mathrm{nuc}/d\log T)$ must be negative immediately below $M_r(T_\mathrm{max})$,
where $\log T$ is decreasing ($d\log T < 0)$, while $\varepsilon_\mathrm{nuc}$ should be increasing
($d\log\varepsilon_\mathrm{nuc} > 0$) with depth (the upper- and middle-left panels in \abb{fig:f3}).
For C burning at $\log T = 8.8 - 8.9$, its energy generation rate can be approximated as
$\varepsilon_\mathrm{nuc} \propto \rho X^2(^{12}\mathrm{C}) T^n$, where $n\approx 40$. Therefore,
\bea
\frac{d\log\varepsilon_\mathrm{nuc}}{d\log T} = \frac{d\log\rho}{d\log T} + 
2 \frac{d\log X(^{12}\mathrm{C})}{d\log T} + n.
\label{eq:dedt}
\eea
From the last equation, it is seen that, in order to get $(d\log\varepsilon_\mathrm{nuc}/d\log T) < 0$, the other two
derivatives have to be negative with relatively large absolute magnitudes to compensate the positive term $n\approx 40$.
A numerical evaluation of the derivatives on the right-hand side of this equation for the profiles
presented in \abb{fig:f3} gives the following estimates: $(d\log\rho/d\log T)\approx -0.7$ and
$(d\log X(^{12}\mathrm{C})/d\log T)\approx -20$. Therefore, we conclude that it is the steep rise of the
$^{12}\mathrm{C}$ mass fraction in the direction toward the center immediately below the point $M_r(T_\mathrm{max})$ 
(the lower-right panel
in \abb{fig:f3}) that secures and maintains the required relative positions of the maxima of 
$\varepsilon_\mathrm{nuc}$ and $\log T$ in the model without extra mixing processes. In such a case, the C flame
propagates all the way down to the center and, as a result, an ONe degenerate core is formed.

\cite{timmes:94} have pointed out that a physically consistent
simulation of the C flame requires a very fine spatial resolution with
mass zones thinner than $\sim 1$ km in the burning region. To comply
with this requirement, \cite{siess:06} used as many as $\sim 50$ grid
points to describe the precursor flame between the bottom of the
C-shell convection zone and the minimum in the luminosity profile
below it. In our simulations of the C-flame propagation, the resolution
in the region of the C-flame precursor is even better
(\abb{fig:f5}). Here, we have more than 100 mass zones
separated by distances less than 1 km. Moreover, our calculated C-flame
speed $V_\mathrm{cond}\approx 1.1\times 10^{-3} \mathrm{cm\, s}^{-1}$ is
in a very good agreement with the value of $V_\mathrm{cond}\sim
(1-4)\times 10^{-3} \mathrm{cm\, s}^{-1}$ interpolated from Tables 1 and
2 of \cite{timmes:94} using the bounding values of $T_9 = 0.76$ and
$\rho = 1.4\times 10^6 \mathrm{g\, cm}^{-3}$ from the model shown in
\abb{fig:f5}. Note that in our model $X(^{12}\mathrm{C})$
increases from $\sim 0.15$ to $\sim 0.35$ in the region of the C-flame
precursor, and \cite{timmes:94} assumed the initial abundances
$X(^{12}\mathrm{C}) = 0.2$ and $X(^{12}\mathrm{C}) = 0.3$ in their
Tables 1 and 2.

\section{The C-flame propagation in the presence of CBM}
\label{sec:ovsh}

In our one-dimensional simulations of CBM, we use a MESA standard option that
allows to take this mixing into account as a diffusion process. The corresponding
diffusion coefficient in radiative layers adjacent to a convective boundary is
\bea 
D_\mathrm{CBM} = D_\mathrm{MLT}(r_0)\exp\left(-\frac{2|r-r_0|}{fH_P}\right),
\label{eq:DCBM}
\eea
where $H_P$ is the pressure scale height and $D_\mathrm{MLT}(r_0)$ is a convective diffusion
coefficient calculated using a mixing-length theory (MLT). 
The MESA {\tt mlt} module assumes that $D_\mathrm{MLT} = \Lambda v_\mathrm{conv}/3$,
where $\Lambda = \alpha H_P$ is the mixing length (we used the value of $\alpha = 2$)
and $v_\mathrm{conv}$ is the convective velocity. The radius $r_0$ is located at the distance $fH_P$ from
the Schwarzschild boundary inside the convective zone. The free parameter $f$ should be
calibrated through observations, or through hydrodynamic simulations.
The exponentially decaying diffusion coefficient (\ref{eq:DCBM}) has succesfully been used
to model CBM at the bottom of the He-shell flash convection zone in AGB stars \citep{herwig:99}
and, more recently, to simulate the interface mixing between a white dwarf and its accreted
H-rich enevelope during thermonuclear runaway of CO nova \citep{denissenkov:13}.
In the first case, using a value of $f\sim 0.008$ produced larger $^{12}$C and $^{16}$O intershell abundances
that were in a better agreement with those observed in the H-deficient post-AGB stars
\citep{werner:06}, while in the second case $f = 0.004$ resulted in the heavy-element enrichment of
nova envelope comparable to those found in multi-dimensional hydrodynamic simulations 
as well as to the spectroscopic measurements of $Z$ in nova ejecta.
For more information on our motivation to use
prescription (\ref{eq:DCBM}) and on the choice of 
appropriate values of $f$ for specific cases, the interested reader is referred to the work of \cite{denissenkov:13}.

\abb{fig:f4} shows the same profiles as \abb{fig:f3},
but for the case when CBM with $f=0.007$ has been included in our
computations. The location of the bottom of the C-shell convective
zone is still tightly bound to that of the $\log T$ maximum, as
explained in the previous section, but now CBM  penetrates
into the convectively stable layers below $M_\mathrm{Sch}$ and homogenizes
the $X(^{12}\mathrm{C})$ distribution, making it almost flat, in these
layers. As a result, the steep rise of $X(^{12}\mathrm{C})$ is moved
away from $M_r(T_\mathrm{max})$, therefore a decrease of $\log T$
immediately below this point cannot be compensated by a sufficiently
strong increase of the $^{12}$C mass fraction necessary to place the
maximum of $\varepsilon_\mathrm{nuc}$ at a sufficient distance in front
of the maximum of $\log T$. In this situation, C ignites on the inner
slope of its profile left from the preceding phase of convective C
burning but, instead of advancing toward the center, the C flame is quenched 
soon after its ignition, when the C abundance is decreasing in
the narrowing convective zone. \abb{fig:f4} gives a snapshot of
such a moment of the C-flame quenching. Although the maximum of
$\varepsilon_\mathrm{nuc}$ is located slightly below
$M_r(T_\mathrm{max})$ in this figure, this is simply because of the fact
that the plasma has not cooled down here yet after the C flame has
passed through it. Therefore, the $\log T$ profile is flatter
immediately below its maximum in this figure than in
\abb{fig:f3}, and for nearly constant $\log T$ and
$X(^{12}\mathrm{C})$ even a small increment of $\log\rho$ with depth
increases $\varepsilon_\mathrm{nuc}$. The described phase of C ignition
followed by the C-flame quenching is repeated many times
(\abb{fig:f6}) until the C abundance  behind the C
flame is reduced to a such low level that its further ignition becomes impossible.
In the end, the C-shell burning with CBM leaves  an unburnt
 CO core ($M_\mathrm{CO}\approx 0.2 \msun$), and the final outcomer is 
 a \emph{hybrid C-O-Ne degenerate core} (\abb{fig:f6}).

\section{Thermohaline mixing in the CO cores of SAGB stars}
\label{sec:th}

Thermohaline mixing develops in situations when the temperature distribution is convectively stable but
the distribution of chemical composition is unstable, provided that heat diffuses faster than the destabilizing chemical 
component. This can occur in the ocean when warm salty water lies 
on top of cold fresh water \citep{stern:60,kunze:03} and in stellar radiative zones when the mean molecular 
weight $\mu$ increases with the radius \citep[e.g.][]{ulrich:72,denissenkov:10}. 
In the oceanic case, thermohaline mixing usually takes the form of vertically elongated
fluid parcels that are called ``salt fingers''.

For the stellar case, a linear stability analysis gives the following estimate for 
a thermohaline diffusion coefficient:
\bea
D_\mathrm{th} = C_\mathrm{th}\,\frac{\nabla_\mu}{\nabla_\mathrm{rad} - \nabla_\mathrm{ad}}K,
\label{eq:Dth}
\eea
where $C_\mathrm{th} = 2\pi^2 a^2$ \citep{denissenkov:10}.
Unfortunately, it is proportional to the square of
the aspect ratio, $a=l/d$, that measures the ratio of the finger's length and diameter.
In his work on the C-flame quenching by thermohaline mixing \cite{siess:09} used the diffusion coefficient
(\ref{eq:Dth}) with $C_\mathrm{th} = 1000$ corresponding to $a\approx 7$ because
that value was known to reproduce the evolutionary decline of the surface C abundance
in low-mass red giants above the bump luminosity \citep{charbonnel:07}.
However, the two- and three-dimensional numerical simulations of the $^3$He-driven thermohaline mixing     
that \cite{charbonnel:07} proposed to lead to the observed C anomaly all resulted in $a < 1$ 
\citep{denissenkov:10,denissenkov:11,traxler:11}.
\cite{denissenkov:10} has explained the low $a$ values by the very low viscosity $\nu$
in the radiative zones of RGB stars which facilitates the development of secondary shear instabilities
that destroy salt fingers \citep{radko:10}.

The linear-theory expression (\ref{eq:Dth}) is valid only for an ideal gas, which is a good approximation for the H-shell burning
in an RGB star. However, the C burning in the cores of SAGB stars occurs under degenerate conditions, in which case
the diffusion coefficient (\ref{eq:Dth}) has to be multiplied by the ratio $(\varphi/\delta)$, where
$\varphi = (\partial\ln\rho/\partial\ln\mu)_{P,T}$ and $\delta = -(\partial\ln\rho/\partial\ln T)_{P,\mu}$
\citep[e.g.,][]{charbonnel:07}.
In the CO cores of SAGB stars, the radiation pressure $P_\mathrm{rad}$ contributes only a few percent to the total one, 
therefore it can be neglected. On the other hand, the pressure of the electron-degenerate gas $P_\mathrm{e}$
exceeds that of the nearly ideal gas of ions by a factor of $\sim 2$ immediately below the C flame 
and by as much as an order of magnitude in the center. When the electron degeneracy increases, the dependence of 
$P_\mathrm{e}$ on $T$ weakens. 
However, $P_\mathrm{e}$ retains a dependence on $\rho$ and $\mu$ similar to that for the ideal gas:
$P_\mathrm{e}\propto (\rho/\mu_\mathrm{e})$, where 
$(1/\mu_\mathrm{e}) = (1/\mu) - \sum_i (X_i/A_i)$, the second term on the right-hand side, the inverse atomic mass averaged over
the distribution of isotopes in mass fractions, representing less than 16\% of the first term in the CO cores. 
Under the circumstances, we expect that $\varphi$ will remain close
to one, while $\delta$ will become very small. Indeed, from the MESA EOS it follows that $\delta$ decreases from            
$\sim 0.8$ to $\sim 0.03$ between the C flame and the center.

\cite{traxler:11} and \cite{denissenkov:11} have shown that 2D and 3D numerical simulations of thermohaline convection for
the RGB case give almost identical estimates of the diffusion coefficient. Therefore, we present here only the results of
2D numerical simulations of thermohaline convection in the CO core of our $9.5 \msun$ solar-metallicity SAGB model star for
parameters extracted from a radiative layer immediately below the C flame. When comparing the simulation parameters 
for the RGB and CO-core cases, the biggest difference is the one between the values of the density ratio 
$R_\rho = (\delta/\varphi)(\nabla - \nabla_\mathrm{ad})/\nabla_\mu$, which are 1700 and 25, respectively.
In the CO-core case, the smaller value of $R_\rho$ is caused by the much larger absolute magnitude of the $\mu$-gradient and
also by $\delta\ll 1$. We have used the same code and resolution $1024\times 1024$ as in the work of
\cite{denissenkov:10}, to which the interested reader is referred for details. 
The results are presented in \abb{fig:f8}, where the red curve in the lower panel plots 
the ratio $D_\mathrm{th}/K$ estimated from
the numerical simulations, while the dashed black and solid blue lines show the same ratio from \eqn{eq:Dth}
calculated for $a = 0.35$ and $a = 1$. 

\abb{fig:f8} shows that the 2D thermohaline diffusion coefficient in
the CO core is very well approximated by the linear-theory one with
$a=0.35$, much like in the RGB case. In our computations of the C-shell convection
 we have used the values of $a=1$ and $a=10$. The first
value is taken a little larger than the one predicted by our numerical
simulations to partly compensate for the decrease of $\delta$ with
depth. The second value is needed, according to our MESA calculations,
to reproduce the evolutionary decline of the surface C abundance above
the bump luminosity in the metal-poor field and globular-cluster red
giants \citep[e.g.,][]{gratton:00,shetrone:10}, assuming that the
observed pattern is produced by the $^3$He-driven thermohaline
convection. Note that the uncertainty factor $C_\mathrm{th} = 1000$ used
by \cite{charbonnel:07} for this purpose is equivalent to $a=7$. When we
insert $a=10$ into \eqn{eq:Dth}, we obtain the results similar to those
of \cite{siess:09}. In this case, as well as in the case of $a=7$, the C
flame stops propagating toward the center by the reason explained in
that paper. The main difference with the CBM case is that the C flame
does not die out soon after being stopped because its life is supported
by the fuel from the underlying CO core supplied by thermohaline
convection, which is not depicted in \abb{fig:f2}. As a result,
only a few percent of C is left unburnt inside the formed ONe core. On
the contrary, in the case with CBM, the C burning leaves a completely
untouched CO core below the final position of the 
quenched C-flame (\abb{fig:f6}). Finally, when we use
the finger aspect ratio $a=1$ that is close to $a=0.35$ derived from our hydrodynamic
simulations\footnote{We take a larger $a$ to partly compensate the factor $\delta^{-2} > 1$ absent in \eqn{eq:Dth}.}, 
thermohaline mixing becomes so inefficient
that it does not interfere with the C-flame propagation toward the center (\abb{fig:f7}).

\section{Accounting for CBM heat transport}
\label{sec:CBM_heat}
In deep stellar interiors, the large convective efficiency, $\Gamma\gg
1$, renders the temperature gradient almost adiabatic in convective
zones, $\nabla\approx\nabla_\mathrm{ad}$, while $\nabla$ remains close
to $\nabla_\mathrm{rad}$ in radiative zones. However, this is true only
if there are no other mixing processes in the radiative zones, or if
such extra mixing is present, but it cannot compete with radiation and
conduction in heat transport. From the mixing length theory \citep[MLT,
e.g.][]{weiss:04}, we estimate $\Gamma = \gamma (D_\mathrm{mix}/K)$,
where $D_\mathrm{mix}$ is a diffusion coefficient describing heat
transport by extra mixing, and $K$ is the thermal diffusivity. For a
given value of $\Gamma$, the MLT gives $\nabla = \zeta\nabla_\mathrm{ad}
+ (1-\zeta)\nabla_\mathrm{rad}$, where
\bea
\zeta = \frac{a_0\Gamma^2}{1+\Gamma(1+a_0\Gamma)}.
\label{eq:zeta}
\eea
In the above equations, $\gamma$ and $a_0$ are factors of the order of one that depend on the geometry of fluid elements.

In the case of thermohaline convection, considered in \kap{sec:th}, our
hydrodynamic simulations give $\Gamma\approx 10^{-3}$ which results in
$\nabla\approx\nabla_\mathrm{rad}$. Therefore, thermohaline convection
should not affect the temperature profile below the C-shell convective
zone. On the contrary, if the prescription (\ref{eq:DCBM}) is used to
also estimate a diffusion coefficient for CBM heat transport then it
will obviously modify $\nabla$ immediately below the C-shell convection
zone. This non-radiative CBM will make $\nabla$ almost adiabatic in the vicinity of the
convective boundary, where $D_\mathrm{CBM}\approx D_\mathrm{MLT}\gg K$,
but it will restore its radiative value at a larger distance, where the
exponential factor reduces $D_\mathrm{CBM}$ below the thermal
diffusivity.

To see how strongly non-radiative CBM may affect the C-flame propagation, we have
computed the evolution of a $7 \msun$ star with $Z=0.01$ from the main
sequence to the C ignition. To produce a more realistic SAGB model, we
have included CBM across the boundaries of the H and He convective cores
and across the bottom of the surface convection zone. For these CBM
processes, we have used \eqn{eq:DCBM} with the value of $f=0.014$ that
is close to the one constrained by fitting the terminal-age main
sequence for a large number of stellar clusters and associations
\citep{herwig:00}. In the case when CBM is included at all evolutionary
phases, both the lower and upper limits for the initial masses of SAGB
stars shift to lower values. To simulate the C-flame propagation with
CBM in the CO core of this SAGB model, we have used the same value of $f
= 0.007$, as in our $9.5 \msun$ model. The CBM heat transport at each
convective boundary has been modeled using the above MLT equations, some
of which are already incorporated into MESA, while others can easily be
added. For the fluid element geometry, we have assumed the following
parameters: $\gamma = 2/3$ and $a_0 = 9/4$ \citep{weiss:04}.

\abb{fig:f9} shows the modification of the temperature
gradient immediately below the C-shell
convection zone produced by the CBM heat transport with the diffusion
coefficient shown in the lower panel. The change of the
temperature stratification occurs only in the immediate vicinity of the
convective boundary because of the short length scale of the exponential
decay of $D_\mathrm{CBM}$. Our computations show that this does not
prevent the C-flame quenching, and again in this case an unburnt CO
core remains in the center (\abb{fig:f10}).

\section{Discussion and conclusions}
\label{sec:concl}

We have obtained the following two main results: 
\begin{enumerate}
\item Convective boundary mixing (CBM) creates conditions in which
the C-flame is inhibited from propagating toward the center in the CO
cores of SAGB stars.
\item Thermohaline mixing driven by the mean molecular weight inversion 
caused by the off-center C burning does not affect the C-flame inward propagation
when one uses a thermohaline diffusion coefficient estimated via
hydrodynamic simulations.
\end{enumerate}

\subsection{Implications for supernova and white dwarf models}
 SAGB stars with off-center C ignition develop hybrid C-O-Ne degenerate cores in their interiors,
rather than ONe cores, because the quenched C flame leaves an unburnt CO core
($M_\mathrm{CO}\approx 0.2 \msun$) in the center, which is surrounded by
a thick ONe zone ($\Delta M_\mathrm{ONe}\approx 0.85 \msun$ by the end
of C burning). The mass of the unburnt core depends not only on the CBM
assumptions, but also on the initial mass as well as on the \czw+\czw\ reaction rate. 
We have carried out a parameter study in which we varied the initial mass, the $f$ value in \eqn{eq:DCBM}, and
the \czw+\czw\ reaction rate to see how those variations affected the masses and chemical structures (pure CO, ONe, or hybrid
C-O-Ne compositions) of the degenerate cores of SAGB stars
(Chen et al., in prep.). It shows that the unburnt CO mass in the hybrid core can be as
large as $\approx 0.45\msun$ for different parameters. 
In addition, in some hybrid models not only the unburnt core but also the outer
C-shell burning layers contain varying amounts of \czw. In
\abb{fig:f10} we show the abundance profiles after
the C-flame has quenched for the case  with the heat-transport
CBM model. The C-shell experienced some intermittent burning which has led to \czw\
abundances in the remaining ONe layer between $2$ and $4\%$ in the
inner region ($<0.7\msun$). Throughout the core the electron fraction remains in
a narrow range $0.4967 < Y_\mem{e} < 0.5$.
 
The hybrid cores may have peculiar properties, compared to CO and ONe
cores, when they evolve subsequently toward a supernova explosion
through one of several possible channels. If an SAGB star with hybrid
core were able to increase the core mass to the Chandrasekhar limit during the
thermal pulse phase (the conditions under which this is possible
for SAGB stars, according to the analysis of models without CBM explored
by \cite{poelarends:08}, apply equally to SAGB stars with hybrid
cores) this could result in a single-star thermonuclear
supernova.

If the remaining C in hybrid cores were indeed sufficient to ignite a
thermonuclear supernova this would have implications for the SN Ia
progenitor models. In the single-degenerate scenario the mass range
of potential progenitors would increase to include those
SAGB stars that generate hybrid cores. This would decrease the
minimum delay time of the appearance of the first SN Ia after a star
formation burst.  It would also provide a larger pool of progenitor
initial masses. Since hybrid cores have larger masses than CO cores,
less mass has to be accreted to reach the Chandrasekhar mass,
compared to model predictions without CBM.

For the double-degenerate scenario we may expect that the ignition of a
C-shell during the merger process may similarly not lead to complete
burning of the core of the primary because in this case the C-flame
may also be quenched. 

If an SAGB star loses its H-rich envelope before its degenerate core
reaches the Chandrasekhar limit, then a hybrid C-O-Ne WD
will result. Due to the different internal composition, the
cooling-down time scale may be different compared to CO and
ONe models. This should be taken into account when using a WD
luminosity function as an age indicator \citep{garcia-berro:97}.

\subsection{Uncertainties}
Our main result seems to be robust, at least in the framework of
one-dimensional simulations. Indeed, the properties of CBM in
environments similar to those at the bottom of the C-shell
convective zone have independently been inferred for the nova case
and for He-shell flash convection in AGB stars, based on both observational data and
results of multi-dimensional hydrodynamic simulations
\citep{werner:06,herwig:06,herwig:07,gehrz:98,glasner:97,herwig:06,
casanova:11}. 

Furthermore, we have considered three different values, namely
$0.004$, $0.007$, and $0.014$, for the $f$ parameter in \eqn{eq:DCBM},
and found the C-flame quenching in all the cases. We have also tried an
experimental two-zone model of CBM that was motivated by hydrodynamic
simulations \citep{herwig:07}. The C-flame behaviour is still very
similar to that obtained with \eqn{eq:DCBM}.

Including CBM during the SAGB progenitor evolution does not change the
results.
We have simulated non-radiative CBM (i.e. accounting for heat transport in
the CBM zone) 
at the boundaries of the H and He
convective cores and at the bottom of the surface convection zone
using \eqn{eq:DCBM} with the value of $f=0.014$ that is close to the
one constrained by fitting the terminal-age main sequence for a large
number of stellar clusters and associations \citep{herwig:00}. In this
case, both the lower and upper limits, $M_\mathrm{up}$ and
$M_\mathrm{mas}$, for the initial masses of SAGB stars shift to lower
values \citep{poelarends:08}. In particular, the evolution of a $7
\msun$ model star with $Z=0.01$ and CBM at the boundaries of its H and
He convective cores ends up with a hybrid C-O-Ne degenerate core very
similar to that in our $9.5 \msun$ model star with $Z=0.02$ in which
such CBM was not taken into account. 

The inclusion of heat transport in the CBM does not have a significant
effect on the pre-SAGB evolution compared to CBM without heat
transport.  Likewise, adding heat-transport to CBM at convective
boundaries does not prevent the C-flame quenching, and also in this
case an unburnt CO core remains inside the SAGB star.

We believe that we have explored all possible effects concerning
species and heat transport by CBM at convective boundaries for two
types of mixing, dynamically induced CBM and secular thermohaline
mixing. However, our investigation is inherently limited by the
assumption of spherical symmetry. The dynamic-timescale mixing
processes discussed here can not be simulated ab-initio in
one-dimensional simulations, but are only approximated with simple
models.  It would be important to explore the implications of hybrid
white dwarfs for the question of how the progenitors of low-mass
supernova, especially those of type Ia, evolve. If the hybrid nature
of super-AGB cores is found to be an important ingredient of promising
progenitor scenarios it would be necessary to further investigate the
physics of the C flame with realistic three-dimensional hydrodynamic
convection and nuclear burning simulations.

\acknowledgements
We thank Bill Merryfield for letting us use his computer code designed
for 2D numerical simulations of salt-fingering convection. FH
appreciates fruitful conversations with Ken Shen that have prompted us
to investigate the effect of heat transport in the CBM region, and that
have informed our conclusions on implications for supernova models. FH
likes to thank Marten van Kerkwijk for an invitation to Toronto and
interesting discussions that have motivated us to complete this project
sooner. This research has been supported by the National Science
Foundation under grants PHY 11-25915 and AST 11-09174. This project was
also supported by JINA (NSF grant PHY 08-22648). FH acknowledges funding
from NSERC through a Discovery Grant. 

\bibliography{ms}

\begin{thebibliography}{46}
\expandafter\ifx\csname natexlab\endcsname\relax\def\natexlab#1{#1}\fi

\bibitem[{{Angulo} {et~al.}(1999){Angulo}, {Arnould}, {Rayet}, \& {et
  al.,}}]{angulo:99}
{Angulo}, C., {Arnould}, M., {Rayet}, M., \& {et al.,}. 1999, Nuclear Physics
  A, 656, 3

\bibitem[{{Casanova} {et~al.}(2011){Casanova}, {Jos{\'e}},
  {Garc{\'{\i}}a-Berro}, {Shore}, \& {Calder}}]{casanova:11}
{Casanova}, J., {Jos{\'e}}, J., {Garc{\'{\i}}a-Berro}, E., {Shore}, S.~N., \&
  {Calder}, A.~C. 2011, \nat, 478, 490

\bibitem[{{Cassisi} {et~al.}(2007){Cassisi}, {Potekhin}, {Pietrinferni},
  {Catelan}, \& {Salaris}}]{cassisi:07}
{Cassisi}, S., {Potekhin}, A.~Y., {Pietrinferni}, A., {Catelan}, M., \&
  {Salaris}, M. 2007, \apj, 661, 1094

\bibitem[{{Caughlan} \& {Fowler}(1988)}]{caughlan:88}
{Caughlan}, G.~R. \& {Fowler}, W.~A. 1988, Atomic Data and Nuclear Data Tables,
  40, 283

\bibitem[{{Charbonnel} \& {Zahn}(2007)}]{charbonnel:07}
{Charbonnel}, C. \& {Zahn}, J.-P. 2007, \aap, 467, L15

\bibitem[{{Denissenkov}(2010)}]{denissenkov:10}
{Denissenkov}, P.~A. 2010, \apj, 723, 563

\bibitem[{{Denissenkov} {et~al.}(2013){Denissenkov}, {Herwig}, {Bildsten}, \&
  {Paxton}}]{denissenkov:13}
{Denissenkov}, P.~A., {Herwig}, F., {Bildsten}, L., \& {Paxton}, B. 2013, \apj,
  762, 8

\bibitem[{{Denissenkov} \& {Merryfield}(2011)}]{denissenkov:11}
{Denissenkov}, P.~A. \& {Merryfield}, W.~J. 2011, \apjl, 727, L8

\bibitem[{{Ferguson} {et~al.}(2005){Ferguson}, {Alexander}, {Allard}, {Barman},
  {Bodnarik}, {Hauschildt}, {Heffner-Wong}, \& {Tamanai}}]{ferguson:05}
{Ferguson}, J.~W., {Alexander}, D.~R., {Allard}, F., {Barman}, T., {Bodnarik},
  J.~G., {Hauschildt}, P.~H., {Heffner-Wong}, A., \& {Tamanai}, A. 2005, \apj,
  623, 585

\bibitem[{{Fynbo} {et~al.}(2005){Fynbo}, {Diget}, {Bergmann}, {et al.,}, \&
  {ISOLDE Collaboration,}}]{fynbo:05}
{Fynbo}, H.~O.~U., {Diget}, C.~A., {Bergmann}, U.~C., {et al.,}, \& {ISOLDE
  Collaboration,}. 2005, \nat, 433, 136

\bibitem[{{Garcia-Berro} \& {Iben}(1994)}]{garcia-berro:94}
{Garcia-Berro}, E. \& {Iben}, I. 1994, \apj, 434, 306

\bibitem[{{Garcia-Berro} {et~al.}(1997){Garcia-Berro}, {Isern}, \&
  {Hernanz}}]{garcia-berro:97}
{Garcia-Berro}, E., {Isern}, J., \& {Hernanz}, M. 1997, \mnras, 289, 973

\bibitem[{{Gehrz} {et~al.}(1998){Gehrz}, {Truran}, {Williams}, \&
  {Starrfield}}]{gehrz:98}
{Gehrz}, R.~D., {Truran}, J.~W., {Williams}, R.~E., \& {Starrfield}, S. 1998,
  \pasp, 110, 3

\bibitem[{{Glasner} {et~al.}(1997){Glasner}, {Livne}, \& {Truran}}]{glasner:97}
{Glasner}, S.~A., {Livne}, E., \& {Truran}, J.~W. 1997, \apj, 475, 754

\bibitem[{{G{\"o}rres} {et~al.}(2000){G{\"o}rres}, {Arlandini}, {Giesen},
  {Heil}, {K{\"a}ppeler}, {Leiste}, {Stech}, \& {Wiescher}}]{gorres:00}
{G{\"o}rres}, J., {Arlandini}, C., {Giesen}, U., {Heil}, M., {K{\"a}ppeler},
  F., {Leiste}, H., {Stech}, E., \& {Wiescher}, M. 2000, \prc, 62, 055801

\bibitem[{{Gratton} {et~al.}(2000){Gratton}, {Sneden}, {Carretta}, \&
  {Bragaglia}}]{gratton:00}
{Gratton}, R.~G., {Sneden}, C., {Carretta}, E., \& {Bragaglia}, A. 2000, \aap,
  354, 169

\bibitem[{{Herwig}(2000)}]{herwig:00}
{Herwig}, F. 2000, \aap, 360, 952

\bibitem[{{Herwig} {et~al.}(1999){Herwig}, {Bl{\"o}cker}, {Langer}, \&
  {Driebe}}]{herwig:99}
{Herwig}, F., {Bl{\"o}cker}, T., {Langer}, N., \& {Driebe}, T. 1999, \aap, 349,
  L5

\bibitem[{{Herwig} {et~al.}(2007){Herwig}, {Freytag}, {Fuchs}, {Hansen},
  {Hueckstaedt}, {Porter}, {Timmes}, \& {Woodward}}]{herwig:07}
{Herwig}, F., {Freytag}, B., {Fuchs}, T., {Hansen}, J.~P., {Hueckstaedt},
  R.~M., {Porter}, D.~H., {Timmes}, F.~X., \& {Woodward}, P.~R. 2007, in
  Astronomical Society of the Pacific Conference Series, Vol. 378, Why Galaxies
  Care About AGB Stars: Their Importance as Actors and Probes, ed.
  F.~{Kerschbaum}, C.~{Charbonnel}, \& R.~F. {Wing}, 43

\bibitem[{{Herwig} {et~al.}(2006){Herwig}, {Freytag}, {Hueckstaedt}, \&
  {Timmes}}]{herwig:06}
{Herwig}, F., {Freytag}, B., {Hueckstaedt}, R.~M., \& {Timmes}, F.~X. 2006,
  \apj, 642, 1057

\bibitem[{{Iglesias} \& {Rogers}(1993)}]{iglesias:93}
{Iglesias}, C.~A. \& {Rogers}, F.~J. 1993, \apj, 412, 752

\bibitem[{{Iglesias} \& {Rogers}(1996)}]{iglesias:96}
---. 1996, \apj, 464, 943

\bibitem[{{Imbriani} {et~al.}(2005){Imbriani}, {Costantini}, {Formicola}, \&
  {et al.,}}]{imbriani:05}
{Imbriani}, G., {Costantini}, H., {Formicola}, A., \& {et al.,}. 2005, European
  Physical Journal A, 25, 455

\bibitem[{{Kunz} {et~al.}(2002){Kunz}, {Fey}, {Jaeger}, {Mayer}, {Hammer},
  {Staudt}, {Harissopulos}, \& {Paradellis}}]{kunz:02}
{Kunz}, R., {Fey}, M., {Jaeger}, M., {Mayer}, A., {Hammer}, J.~W., {Staudt},
  G., {Harissopulos}, S., \& {Paradellis}, T. 2002, \apj, 567, 643

\bibitem[{{Kunze}(2003)}]{kunze:03}
{Kunze}, E. 2003, Progress in Oceanography, 56, 399

\bibitem[{{Miller Bertolami} {et~al.}(2006){Miller Bertolami}, {Althaus},
  {Serenelli}, \& {Panei}}]{miller-bertolami:06}
{Miller Bertolami}, M.~M., {Althaus}, L.~G., {Serenelli}, A.~M., \& {Panei},
  J.~A. 2006, \aap, 449, 313

\bibitem[{{Miyaji} {et~al.}(1980){Miyaji}, {Nomoto}, {Yokoi}, \&
  {Sugimoto}}]{miyaji:80}
{Miyaji}, S., {Nomoto}, K., {Yokoi}, K., \& {Sugimoto}, D. 1980, \pasj, 32, 303

\bibitem[{{Paxton} {et~al.}(2011){Paxton}, {Bildsten}, {Dotter}, {Herwig},
  {Lesaffre}, \& {Timmes}}]{paxton:11}
{Paxton}, B., {Bildsten}, L., {Dotter}, A., {Herwig}, F., {Lesaffre}, P., \&
  {Timmes}, F. 2011, \apjs, 192, 3

\bibitem[{{Paxton} {et~al.}(2013){Paxton}, {Cantiello}, {Arras}, {Bildsten},
  {Brown}, {Dotter}, {Mankovich}, {Montgomery}, {Stello}, {Timmes}, \&
  {Townsend}}]{paxton:13}
{Paxton}, B., {Cantiello}, M., {Arras}, P., {Bildsten}, L., {Brown}, E.~F.,
  {Dotter}, A., {Mankovich}, C., {Montgomery}, M.~H., {Stello}, D., {Timmes},
  F.~X., \& {Townsend}, R. 2013, ArXiv e-prints

\bibitem[{{Poelarends} {et~al.}(2008){Poelarends}, {Herwig}, {Langer}, \&
  {Heger}}]{poelarends:08}
{Poelarends}, A.~J.~T., {Herwig}, F., {Langer}, N., \& {Heger}, A. 2008, \apj,
  675, 614

\bibitem[{{Potekhin} \& {Chabrier}(2010)}]{potekhin:10}
{Potekhin}, A.~Y. \& {Chabrier}, G. 2010, Contributions to Plasma Physics, 50,
  82

\bibitem[{{Pumo} {et~al.}(2009){Pumo}, {Turatto}, {Botticella}, {Pastorello},
  {Valenti}, {Zampieri}, {Benetti}, {Cappellaro}, \& {Patat}}]{pumo:09}
{Pumo}, M.~L., {Turatto}, M., {Botticella}, M.~T., {Pastorello}, A., {Valenti},
  S., {Zampieri}, L., {Benetti}, S., {Cappellaro}, E., \& {Patat}, F. 2009,
  \apjl, 705, L138

\bibitem[{{Radko}(2010)}]{radko:10}
{Radko}, T. 2010, Journal of Fluid Mechanics, 645, 121

\bibitem[{{Rogers} \& {Nayfonov}(2002)}]{rogers:02}
{Rogers}, F.~J. \& {Nayfonov}, A. 2002, \apj, 576, 1064

\bibitem[{{Saumon} {et~al.}(1995){Saumon}, {Chabrier}, \& {van
  Horn}}]{saumon:95}
{Saumon}, D., {Chabrier}, G., \& {van Horn}, H.~M. 1995, \apjs, 99, 713

\bibitem[{{Shetrone} {et~al.}(2010){Shetrone}, {Martell}, {Wilkerson}, {Adams},
  {Siegel}, {Smith}, \& {Bond}}]{shetrone:10}
{Shetrone}, M., {Martell}, S.~L., {Wilkerson}, R., {Adams}, J., {Siegel},
  M.~H., {Smith}, G.~H., \& {Bond}, H.~E. 2010, \aj, 140, 1119

\bibitem[{{Siess}(2006)}]{siess:06}
{Siess}, L. 2006, \aap, 448, 717

\bibitem[{{Siess}(2009)}]{siess:09}
---. 2009, \aap, 497, 463

\bibitem[{{Stern}(1960)}]{stern:60}
{Stern}, M.~E. 1960, Tellus, 12, 172

\bibitem[{{Timmes} \& {Swesty}(2000)}]{timmes:00}
{Timmes}, F.~X. \& {Swesty}, F.~D. 2000, \apjs, 126, 501

\bibitem[{{Timmes} {et~al.}(1994){Timmes}, {Woosley}, \& {Taam}}]{timmes:94}
{Timmes}, F.~X., {Woosley}, S.~E., \& {Taam}, R.~E. 1994, \apj, 420, 348

\bibitem[{{Traxler} {et~al.}(2011){Traxler}, {Garaud}, \&
  {Stellmach}}]{traxler:11}
{Traxler}, A., {Garaud}, P., \& {Stellmach}, S. 2011, \apjl, 728, L29

\bibitem[{{Ulrich}(1972)}]{ulrich:72}
{Ulrich}, R.~K. 1972, \apj, 172, 165

\bibitem[{{Weiss} \& {Ferguson}(2009)}]{weiss:09}
{Weiss}, A. \& {Ferguson}, J.~W. 2009, \aap, 508, 1343

\bibitem[{{Weiss} {et~al.}(2004){Weiss}, {Hillebrandt}, {Thomas}, \&
  {Ritter}}]{weiss:04}
{Weiss}, A., {Hillebrandt}, W., {Thomas}, H.-C., \& {Ritter}, H. 2004, {Cox and
  Giuli's Principles of Stellar Structure}

\bibitem[{{Werner} \& {Herwig}(2006)}]{werner:06}
{Werner}, K. \& {Herwig}, F. 2006, \pasp, 118, 183

\end{thebibliography}

\newpage

\begin{figure}
\epsfxsize=10cm
\epsffile [10 150 350 650] {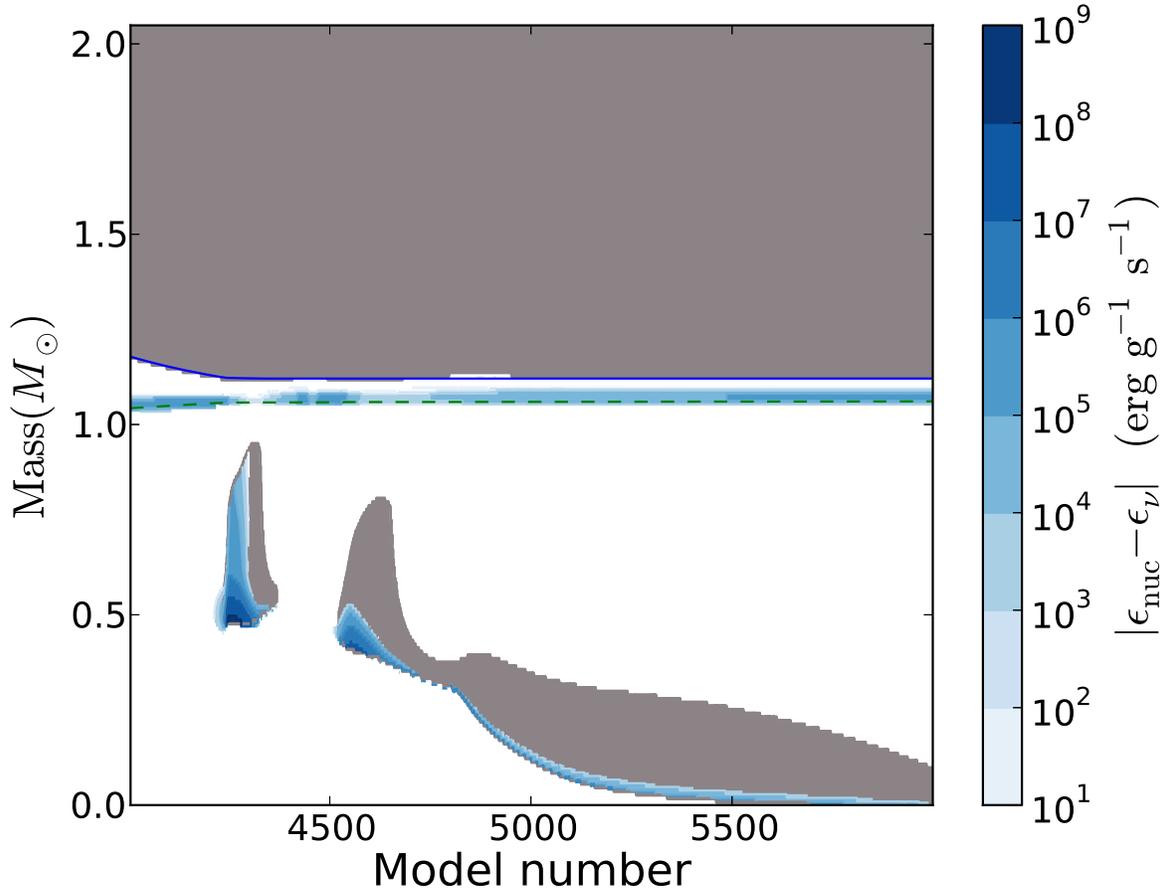}
\caption{A Kippenhahn diagram for the phase of C burning in the core of
our $9.5\msun$ SAGB star model with the near solar metallicity $Z=0.02$
in the absence of extra mixing, such as thermohaline convection or
convective boundary mixing. The C flash is followed by stationary C
burning propagating all the way down to the center. The uniform grey
areas are convective zones. The different shades of blue color map the
nuclear energy generation rate.}
\label{fig:f1}
\end{figure}

\begin{figure}
\epsfxsize=10cm
\epsffile [10 150 350 650] {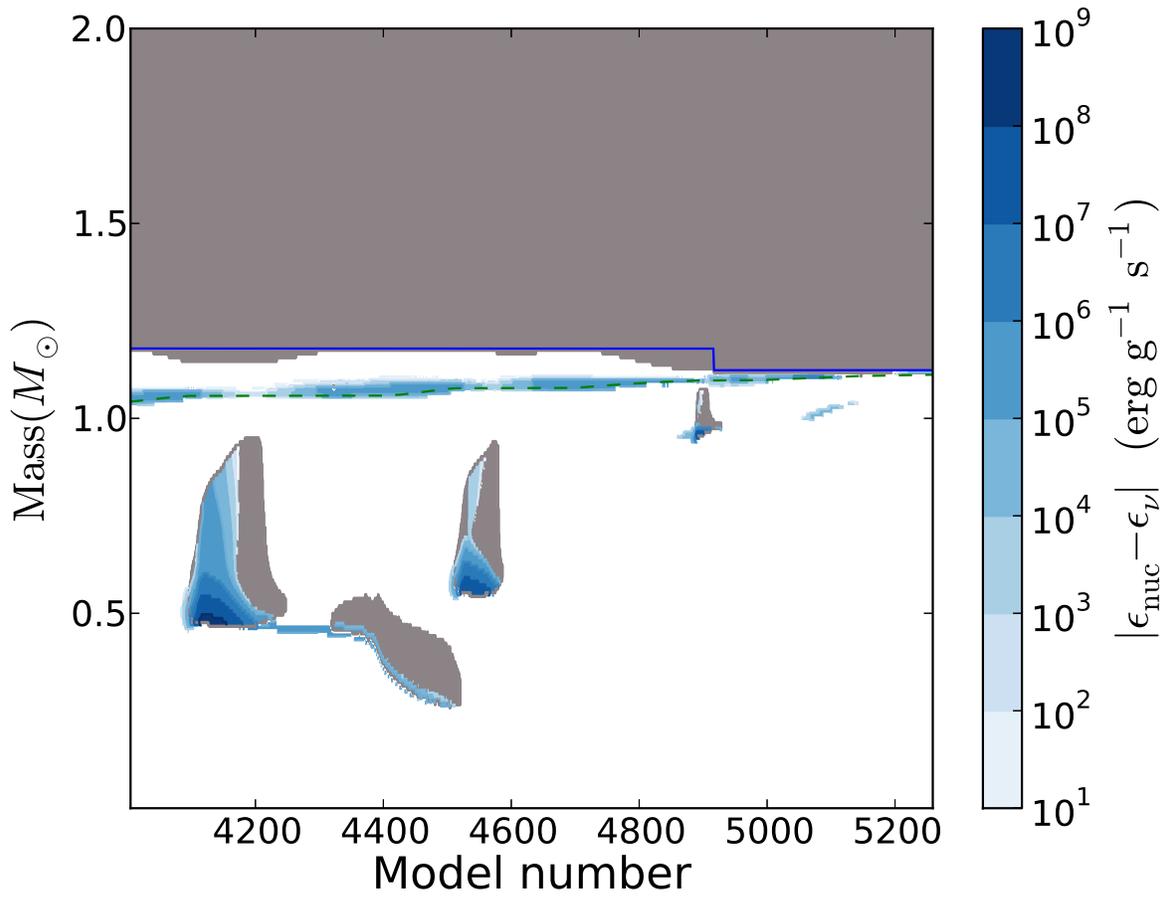}
\caption{Same as in \abb{fig:f1}, but in the presence of thermohaline mixing with the salt-finger aspect ratio $a=10$. 
In this case, the C-flame fails to propagate to the center leaving below a relatively large unburnt CO core.}
\label{fig:f2}
\end{figure}

\begin{figure}
\epsfxsize=10cm
\epsffile [80 210 380 695] {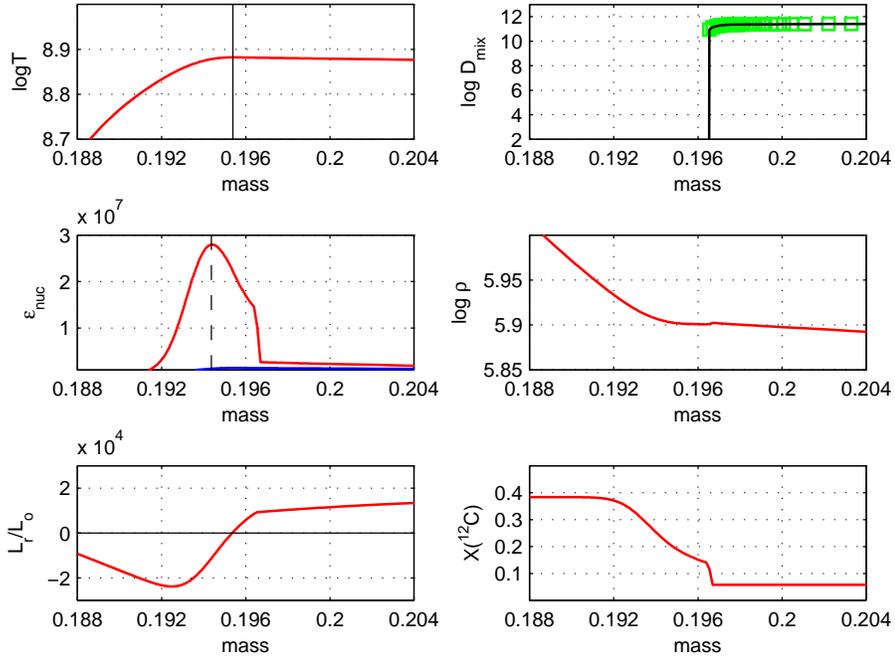}
\caption{The temperature ($\log T$), diffusion coefficient ($\log
D_\mathrm{mix}$), nuclear energy generation rate
($\varepsilon_\mathrm{nuc}$), density ($\log\rho$), luminosity
($L_r/L_\odot$), and $^{12}$C mass fraction ($X(^{12}\mathrm{C}$))
profiles, as functions of the Lagrangian mass coordinate ($\mathrm{mass}
= M_r/\msun$), in the vicinity of the bottom of the C-flame convective
zone in the model 4900 from \abb{fig:f1}. The vertical solid and
dashed lines in the upper- and middle-left panels show the maxima of the
corresponding curves. The green squares in the upper-right panel give
the convective diffusion coefficient, while the solid black curve adds
up diffusion coefficients from all mixing processes. The blue curve at
the bottom of the middle-left plot shows the neutrino energy loss rate.}
\label{fig:f3}
\end{figure}

\begin{figure}
\epsfxsize=10cm
\epsffile [80 210 380 695] {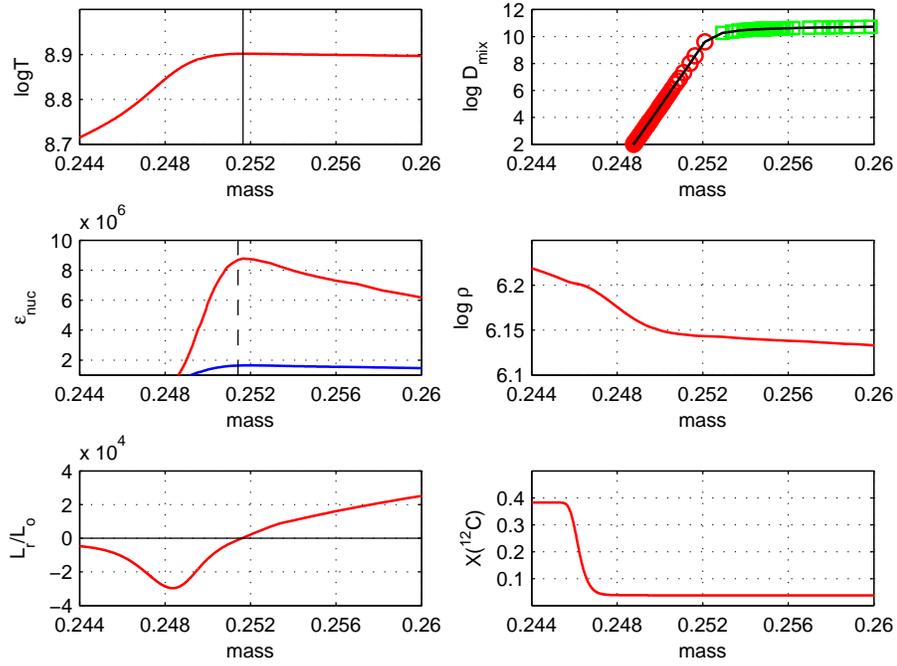}
\caption{Same as in \abb{fig:f3}, but for the case when CBM with $f=0.007$ is taken into account for the model 5073 
from \abb{fig:f6}. The red circles in the upper-right panel show the diffusion coefficient for CBM.}
\label{fig:f4}
\end{figure}

\begin{figure}
\epsfxsize=10cm
\epsffile [10 150 350 650] {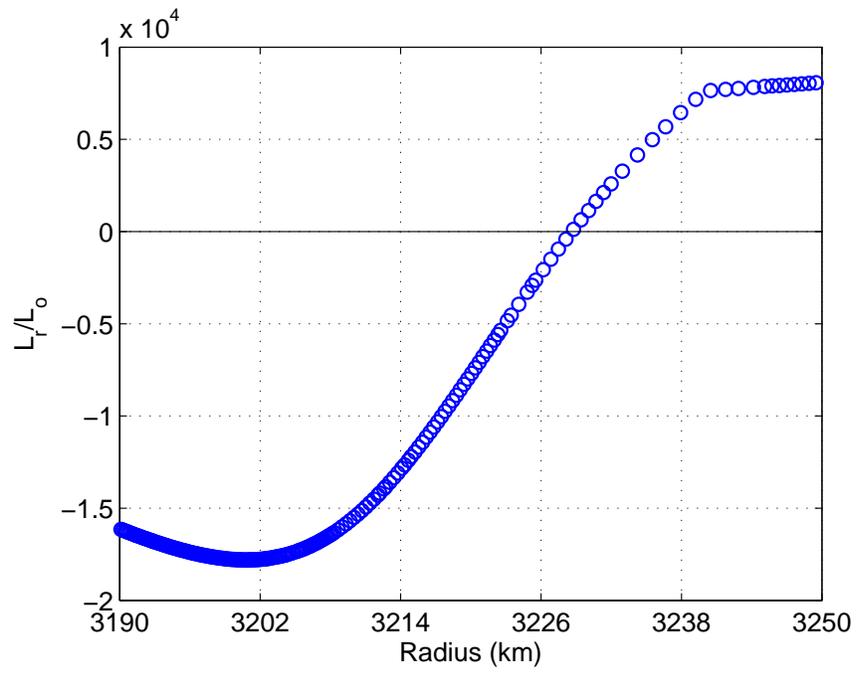}
\caption{A typical mass zoning in the region of the C-flame precursor in our simulations of the C-flame propagation.
        }
\label{fig:f5}
\end{figure}

\begin{figure}
\epsfxsize=10cm
\epsffile [10 150 350 650] {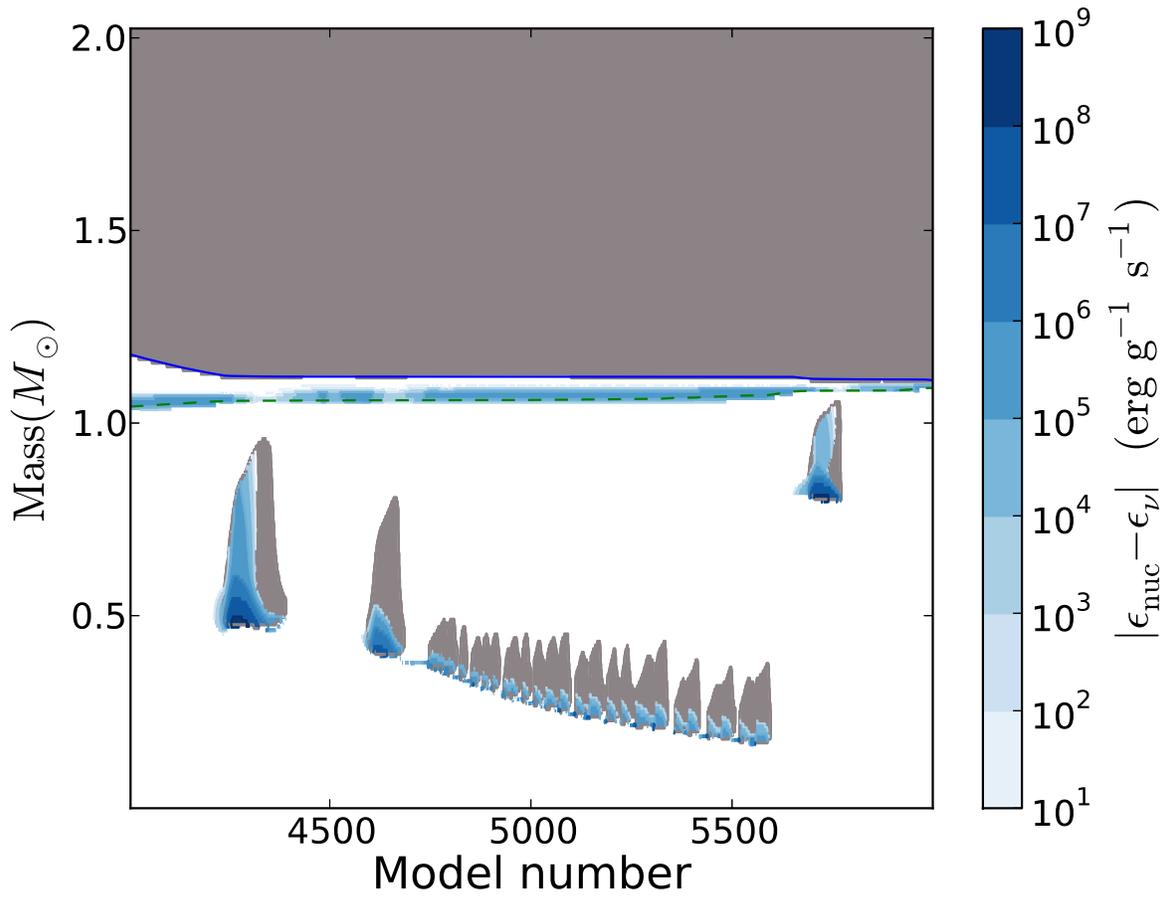}
\caption{Same as in \abb{fig:f2}, but in the presence of CBM with $f=0.007$.
In this case, the C-flame also fails to propagate to the center.}
\label{fig:f6}
\end{figure}

\begin{figure}
\epsfxsize=10cm
\epsffile [10 150 350 650] {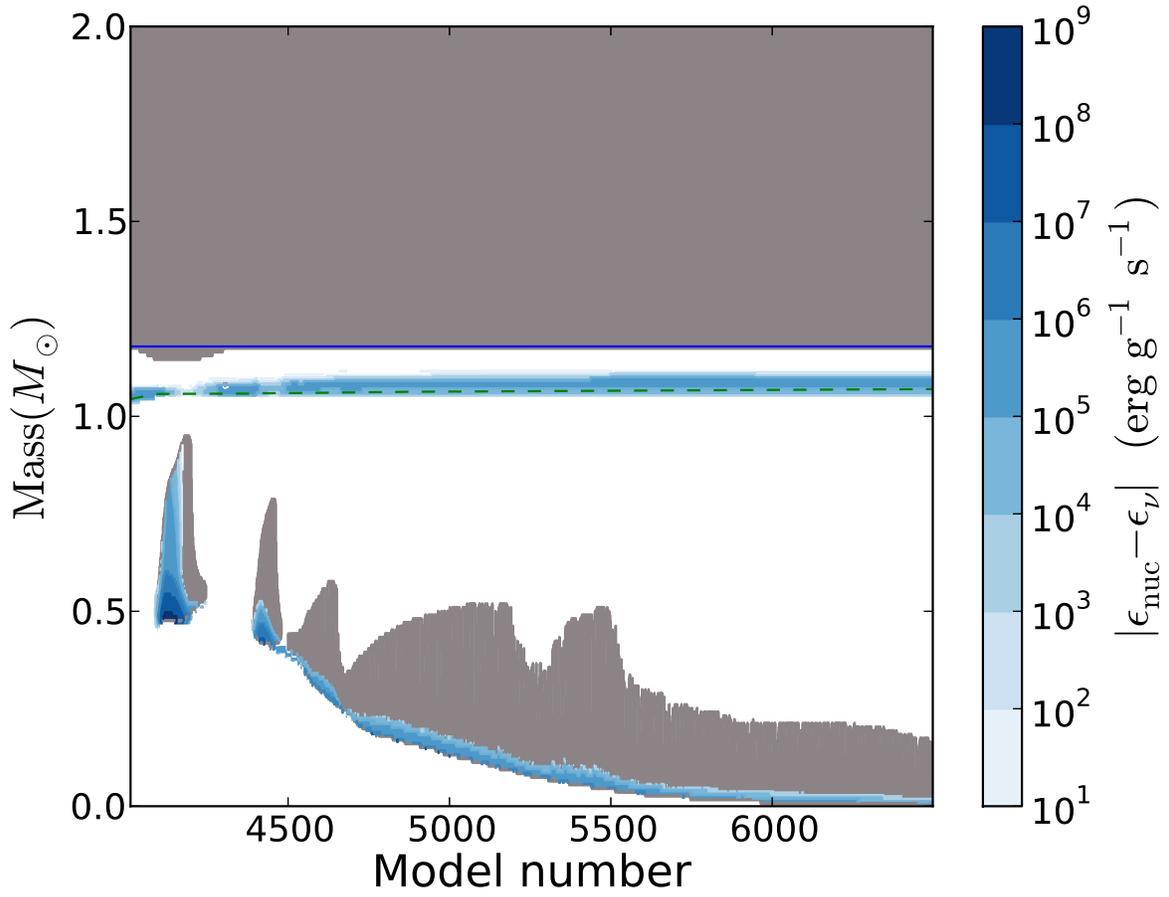}
\caption{Same as in \abb{fig:f2}, but for the finger aspect ratio $a=1$.
In this case, the C-flame propagates to the center.}
\label{fig:f7}
\end{figure}

\begin{figure}
\epsfxsize=10cm
\epsffile [10 150 350 650] {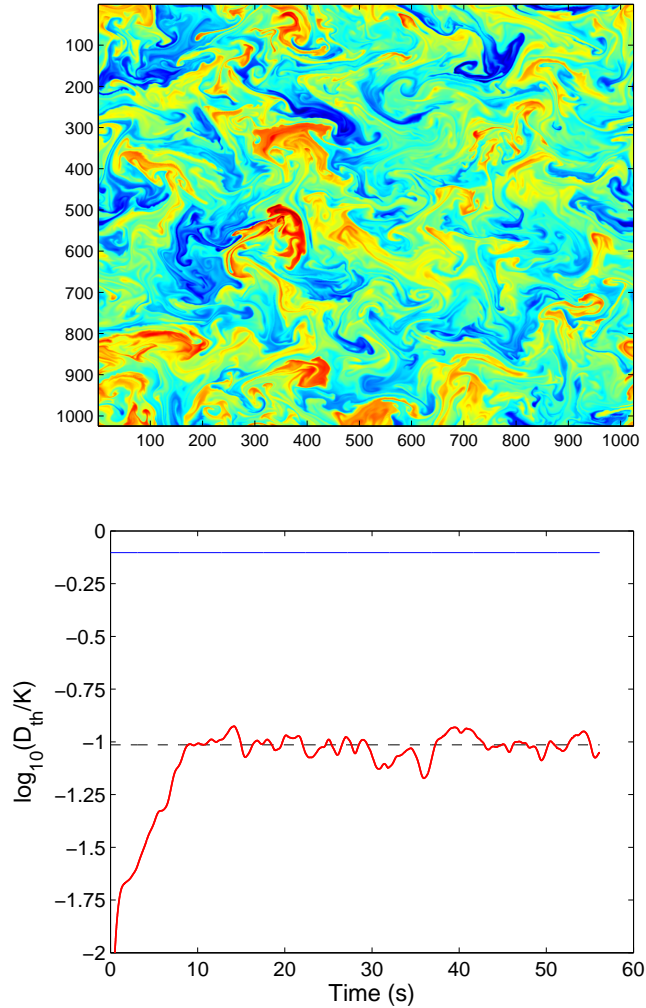}
\caption{The results of our 2D numerical simulations of thermohaline convection in a zone immediately below
the bottom of the C-shell convection. Upper panel: the developed fluid motion does not show vertically elongated
structures because the growing salt fingers are destroyed by the secondary instability 
(the color changing from red to blue corresponds to
the increasing mean molecular weight).
Lower panel: The red curve shows the result of numerical simulations, while 
the dashed black and solid blue lines represent the linear-theory 
approximation (\ref{eq:Dth}) for the finger aspect ratios $a=0.35$ and $a=1$.}
\label{fig:f8}
\end{figure}

\begin{figure}
\epsfxsize=10cm
\epsffile [10 150 350 650] {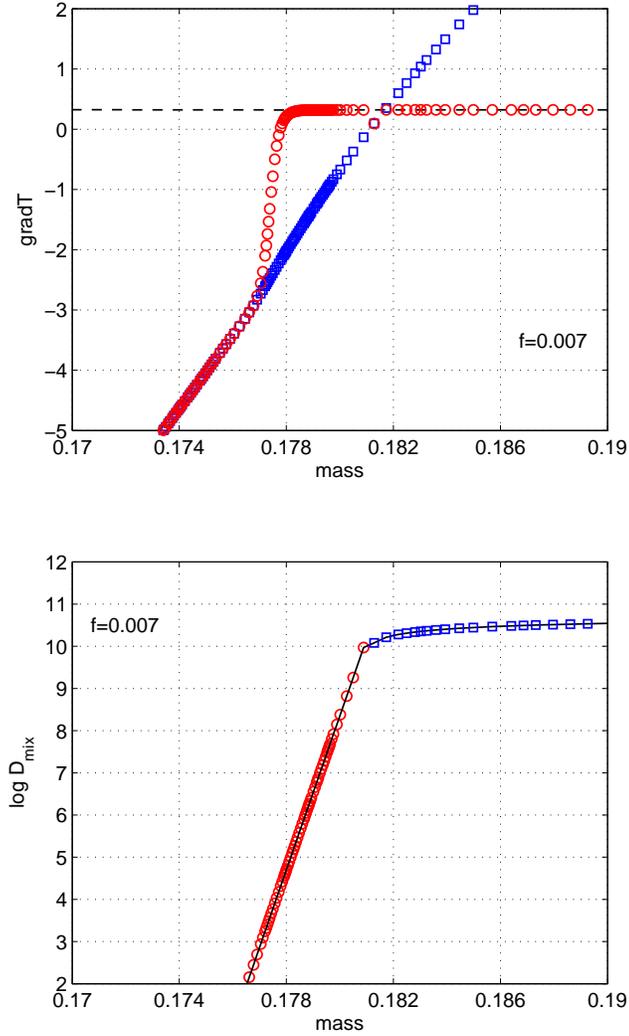}
\caption{Upper panel: blue squares show the radiative temperature
gradient, and red circles show the true temperature gradient modified by
CBM heat transport in the vicinity of the convective boundary (a
step-like profile). The dashed line shows $\nabla_\mem{ad}$. 
Lower panel: blue squares show the MLT diffusion
coefficient in the C-shell convective zone, and red circles show the CBM
diffusion coefficient (\ref{eq:DCBM}) for $f=0.007$. The latter has been
used to modify the true temperature gradient in the upper panel, as
described in text.}
\label{fig:f9}
\end{figure}

\begin{figure}
\epsfxsize=10cm
\epsffile [10 150 350 650] {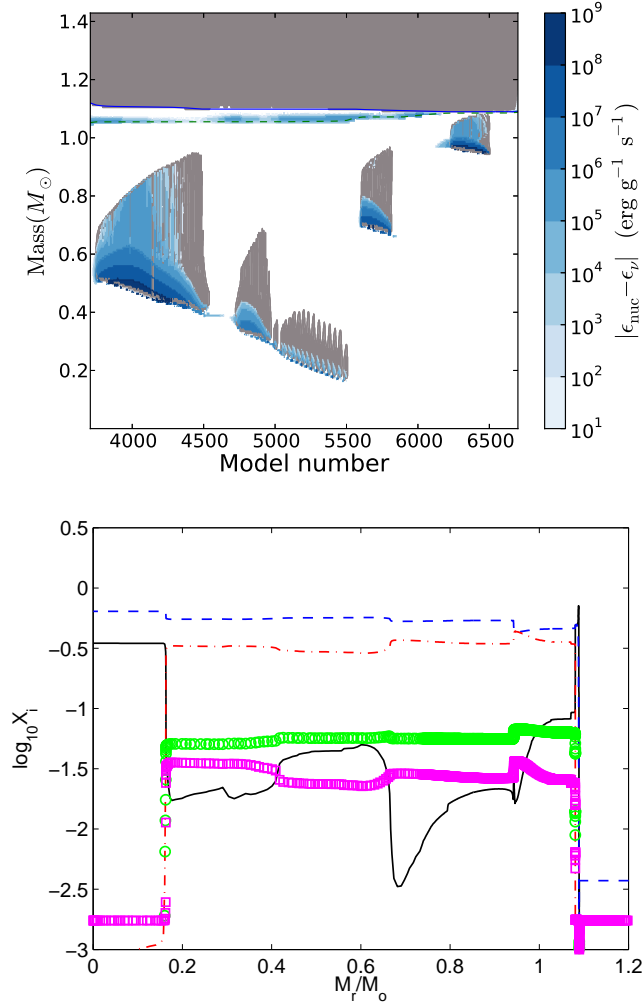}
\caption{Top panel: Same as in \abb{fig:f6}, but for the $7
\msun$ star with $Z=0.01$. Both CBM and its heat transport have been
taken into account using the MLT prescription (see text). We have used
$f=0.014$ during the evolution preceding the C ignition, and $f =
0.007$ at the boundaries of the C-shell convection zone. In this case,
the C-flame is quenched before reaching the center. Bottom panel:
Abundance mass fraction profiles in the final model of this sequence
(solid black: \czw; dashed blue: \ose; dot-dashed red: \nezwa; green
circles: \nadr; magenta squares: \mgvi.}
\label{fig:f10}
\end{figure}

\end{document}